\documentclass[letter, traditabstract, longauth]{aa}
\usepackage{graphicx, natbib, txfonts}
\bibpunct{(}{)}{;}{a}{}{,}

\begin{document}

\title{SPIRE imaging of M\,82: cool dust in the wind and tidal
streams\thanks{Herschel is an ESA space observatory with science instruments provided
by European-led consortia, and important participation of NASA.}}

\author{H. Roussel\inst{\ref{inst9}},
C. D. Wilson\inst{\ref{inst20}},
L. Vigroux\inst{\ref{inst9}},
K. G. Isaak\inst{\ref{inst2},\ref{inst23}},
M. Sauvage\inst{\ref{inst11}},
S. C. Madden\inst{\ref{inst11}},
R. Auld\inst{\ref{inst2}},
M. Baes\inst{\ref{inst5}},
M. J. Barlow\inst{\ref{inst6}},
G. J. Bendo\inst{\ref{inst3}},
J. J. Bock\inst{\ref{inst7}},
A. Boselli\inst{\ref{inst1}},
M. Bradford\inst{\ref{inst7}},
V. Buat\inst{\ref{inst1}},
N. Castro-Rodriguez\inst{\ref{inst8}},
P. Chanial\inst{\ref{inst11}},
S. Charlot\inst{\ref{inst9}},
L. Ciesla\inst{\ref{inst1}},
D. L. Clements\inst{\ref{inst3}},
A. Cooray\inst{\ref{inst25}},
D. Cormier\inst{\ref{inst11}},
L. Cortese\inst{\ref{inst2}},
J. I. Davies\inst{\ref{inst2}},
E. Dwek\inst{\ref{inst10}},
S. A. Eales\inst{\ref{inst2}},
D. Elbaz\inst{\ref{inst11}},
M. Galametz\inst{\ref{inst11}},
F. Galliano\inst{\ref{inst11}},
W. K. Gear\inst{\ref{inst2}},
J. Glenn\inst{\ref{inst13}},
H. L. Gomez\inst{\ref{inst2}},
M. Griffin\inst{\ref{inst2}},
S. Hony\inst{\ref{inst11}},
L. R. Levenson\inst{\ref{inst7}},
N. Lu\inst{\ref{inst7}},
B. O'Halloran\inst{\ref{inst3}},
K. Okumura\inst{\ref{inst11}},
S. Oliver\inst{\ref{inst14}},
M. J. Page\inst{\ref{inst15}},
P. Panuzzo\inst{\ref{inst11}},
A. Papageorgiou\inst{\ref{inst2}},
T. J. Parkin\inst{\ref{inst20}},
I. Perez-Fournon\inst{\ref{inst8}},
M. Pohlen\inst{\ref{inst2}},
N. Rangwala\inst{\ref{inst13}},
E. E. Rigby\inst{\ref{inst4}},
A. Rykala\inst{\ref{inst2}},
N. Sacchi\inst{\ref{inst17}},
B. Schulz\inst{\ref{inst16}},
M. R. P. Schirm\inst{\ref{inst20}},
M. W. L. Smith\inst{\ref{inst2}},
L. Spinoglio\inst{\ref{inst17}},
J. A. Stevens\inst{\ref{inst18}},
S. Srinivasan\inst{\ref{inst9}},
M. Symeonidis\inst{\ref{inst15}},
M. Trichas\inst{\ref{inst3}},
M. Vaccari\inst{\ref{inst19}},
H. Wozniak\inst{\ref{inst21}},
G. S. Wright\inst{\ref{inst24}},
W. W. Zeilinger\inst{\ref{inst22}}
}

\institute{
Institut d'Astrophysique de Paris, UMR7095 CNRS, Universit\'e Pierre \& Marie Curie, 98 bis Boulevard Arago, F-75014 Paris, France\label{inst9}
   \email{roussel@iap.fr}
\and Dept. of Physics \& Astronomy, McMaster University, Hamilton, Ontario, L8S 4M1, Canada\label{inst20}
\and School of Physics \& Astronomy, Cardiff University, Queens Buildings The Parade, Cardiff CF24 3AA, UK\label{inst2}
\and ESA Astrophysics Missions Division, ESTEC, PO Box 299, 2200 AG Noordwijk, The Netherlands\label{inst23} 
\and CEA, Laboratoire AIM, Irfu/SAp, Orme des Merisiers, F-91191 Gif-sur-Yvette, France\label{inst11}
\and Sterrenkundig Observatorium, Universiteit Gent, Krijgslaan 281 S9, B-9000 Gent, Belgium\label{inst5}
\and Dept. of Physics \& Astronomy, University College London, Gower Street, London WC1E 6BT, UK\label{inst6}
\and Astrophysics Group, Imperial College, Blackett Laboratory, Prince Consort Road, London SW7 2AZ, UK\label{inst3}
\and Jet Propulsion Laboratory, Pasadena, CA 91109, United States;  
     Dept. of Astronomy, California Institute of Technology, Pasadena, CA 91125, USA\label{inst7}
\and Laboratoire d'Astrophysique de Marseille, UMR6110 CNRS, 38 rue F. Joliot-Curie, F-13388 Marseille France\label{inst1}
\and Instituto de Astrof\'isica de Canarias, v\'ia L\'actea S/N, E-38200 La Laguna, Spain\label{inst8}
\and Dept. of Physics \& Astronomy, University of California, Irvine, CA 92697, USA\label{inst25}
\and Observational  Cosmology Lab, Code 665, NASA Goddard Space Flight Center Greenbelt, MD 20771, USA\label{inst10}
\and Dept. of Astrophysical \& Planetary Sciences, CASA CB-389, University of Colorado, Boulder, CO 80309, USA\label{inst13}
\and Astronomy Centre, Dept. of Physics and Astronomy, University of Sussex, UK\label{inst14}
\and Mullard Space Science Laboratory, University College London, Holmbury St Mary, Dorking, Surrey RH5 6NT, UK\label{inst15}
\and School of Physics \& Astronomy, University of Nottingham, University Park, Nottingham NG7 2RD, UK\label{inst4}
\and Istituto di Fisica dello Spazio Interplanetario, INAF, Via del Fosso del Cavaliere 100, I-00133 Roma, Italy\label{inst17}
\and Infrared Processing \& Analysis Center, California Institute of Technology, Mail Code 100-22, 770 South Wilson Av, Pasadena, CA 91125, USA\label{inst16}
\and Centre for Astrophysics Research, Science \& Technology Research Centre, University of Hertfordshire, College Lane, Herts AL10 9AB, UK\label{inst18}
\and University of Padova, Dept. of Astronomy, Vicolo Osservatorio 3, I-35122 Padova, Italy\label{inst19}
\and Observatoire Astronomique de Strasbourg, UMR 7550 Universit\'e de Strasbourg - CNRS, 11, rue de l'Universit\'e, F-67000 Strasbourg\label{inst21}
\and UK Astronomy Technology Center, Royal Observatory Edinburgh, Edinburgh, EH9 3HJ, UK\label{inst24} 
\and Institut f\"ur Astronomie, Universit\"at Wien, T\"urkenschanzstr. 17, A-1180 Wien, Austria\label{inst22}
}

\abstract{
M\,82 is a unique representative of a whole class of galaxies, starbursts with superwinds,
in the Very Nearby Galaxy Survey with Herschel. In addition, its interaction with the M\,81
group has stripped a significant portion of its interstellar medium from its disk.
SPIRE maps now afford better characterization of the far-infrared emission from cool dust outside
the disk, and sketch a far more complete picture of its mass distribution and energetics
than previously possible. They show emission coincident in projection with the starburst
wind and in a large halo, much more extended than the PAH band emission seen with Spitzer.
Some complex substructures coincide with the brightest PAH filaments, and others with
tidal streams seen in atomic hydrogen.
We subtract the far-infrared emission of the starburst and underlying disk from the maps,
and derive spatially-resolved far-infrared colors for the wind and halo. We interpret the
results in terms of dust mass, dust temperature, and global physical conditions.
In particular, we examine variations in the dust physical properties as a function of distance
from the center and the wind polar axis, and conclude that more than two thirds
of the extraplanar dust has been removed by tidal interaction, and not entrained by the
starburst wind.}
 

\authorrunning{Roussel et al.}

\date{Received 30/03/2010 / Accepted 13/04/2010}

\maketitle

\section{Introduction}

\object{M\,82} is the closest specimen of starbursts with superwinds, at a distance of only
$(3.9 \pm 0.3)$\,Mpc \citep{Sakai99}. Its almost edge-on orientation is ideal for
the study of the superwind, and more generally the interaction of galaxies with the
intergalactic medium and the role of the gas and dust dynamics in the self-regulation
of starbursts.

The exceptional star formation activity of M\,82 is thought to be caused by its tidal
interaction with the M\,81 galaxy group, redistributing the gas within the disk of M\,82
toward the central regions where it accumulates, and expelling atomic gas in giant streams
connecting the disk to the intragalactic medium \citep{Yun94}.
M\,82 is an evolved starburst, in the sense that its activity has been sustained
(episodically) over much of its lifetime.
Coupled stellar population synthesis and photoionization models reveal
two major episodes of star formation over the past 10\,Myr \citep{Forster03}.
Numerous young supernova remnants have been detected \citep{Muxlow94}.
Another peak of star formation about 150\,Myr ago is inferred from the age-dating
of stellar clusters \citep{Konstantopoulos09}, and 200 super star clusters,
signposts of starbursts, have been found in M\,82 \citep{Melo05}, spanning a wide
age interval from 60\,Myr to several Gyr \citep{Gallagher99, Grijs01}.

\cite{Ward03} found that more than half the total molecular gas of the central regions
is warmer than 50\,K, and \cite{Mao00} found that CO transitions up to $J=7-6$
trace diffuse gas abundantly illuminated by UV radiation. \cite{Panuzzo10}
modeled CO transitions from $J=4-3$ to $J=13-12$ in the central 43$^{\prime\prime}$,
observed with the SPIRE FTS spectrometer, and they suggest that turbulent heating caused
by stellar winds and supernovae may be the dominant excitation process for the highest
transitions, tracing gas at about 500\,K. The molecular gas properties thus reinforce
the notion that M\,82 is an evolved starburst, which has been active for long enough
to completely disrupt the cold component of the interstellar medium.

The mechanical energy simultaneously released by a large collection of massive stars
and supernovae is able to drive a superwind detected in both X-rays emitted by shocked gas
\citep{Strickland97} and optical lines excited by a combination of photoionization
and shock ionization \citep{Shopbell98}. Optical emission around the wind also partly
issues from a reflection nebula \citep{Scarrott91}, indicating the presence of dust.
Abundant molecular gas is associated with both the superwind (although probably not
spatially coincident but rather outside the bicylindrical outflow) and tidal
streams \citep{Walter02}, and PAH emission is found in a very extended halo
\citep{Engelbracht06}.

This paper studies the spatially-resolved properties of the cool dust detected by SPIRE.
Since PACS photometric observations had not been scheduled at the time of writing, and
Spitzer MIPS images are saturated, we sample the far-infrared spectral energy distribution
only longward of 250\,$\mu$m, where it can be safely assumed that the emission is dominated
by grains in thermal equilibrium. SCUBA maps have higher angular resolutions than SPIRE maps;
their sensitivities are however too low to reliably detect any emission outside the disk
of M\,82, even when all archival data are combined \citep{Leeuw09}. In contrast, bright
filaments and diffuse emission are detected out to very large distances with SPIRE.
We thus focus on the dust content and physical properties outside the main body
of M\,82, where SPIRE maps provide truly unique information.

\section{SPIRE observations, data reduction, and analysis}

M\,82 was observed with the SPIRE photometer arrays onboard Herschel \citep{Pilbratt10}
during the science demonstration phase. The SPIRE instrument, its in-orbit
performance, and scientific capabilities are described by \cite{Griffin10}.
Its resolving power enables a linear resolution of 350 to 700\,pc
in M\,82, from 250 to 500\,$\mu$m (1\,kpc corresponds to $53^{\prime\prime}$).

SPIRE maps of $22^{\prime} \times 22^{\prime}$ were acquired in nominal scan mode, with
a total of 4 scans. The data were processed with {\small HIPE} 3.0 up to level 1, as described
by \cite{Pohlen10}, except without the thermal drift correction.
The data were then input to the Scanamorphos software (Roussel, in preparation),
which produces maps after subtracting the thermal drift and the low-frequency noise of
each bolometer, purely by using the available redundancy.
The resulting maps have $\sim 30$\% smaller sky standard deviations and $\sim 3$\%
larger beam areas.

We also derived SPIRE point response functions (PRF) \footnote{This term designates
the result of sampling the point spread function (PSF) of the optics by the combination
of the observation pattern and the map pixel grid: it is an empirical realization of the PSF.}
from a nominal 8-scan observation of Uranus, with a field of view of
$20^{\prime} \times 20^{\prime}$, that we processed in the same way and with the
same orientation along the first scan\footnote{files available
at {\it http://www2.iap.fr/users/roussel/spirebeams/}}. This observation is deeper than
that of Neptune -- made in bright-source mode -- used for the official PRF release, and
thus allows detection of the secondary lobes and diffraction spikes.
Uranus is slightly more extended than Neptune ($3.53^{\prime\prime}$ at the time of
observation\footnote{using the {\small HORIZONS} ephemeris service accessible at
{\it http://ssd.jpl.nasa.gov/horizons.cgi}}, as opposed to $2.26^{\prime\prime}$),
but suitable for sources larger than 70\,pc in M\,82. To create the PRF, all the emission
outside the beam pattern was masked out, and the few remaining background sources were
approximated by Gaussians and subtracted.

All maps have a pixel size of $4.5^{\prime\prime}$ at 250\,$\mu$m, $6.25^{\prime\prime}$ at
350\,$\mu$m, and $9^{\prime\prime}$ at 500\,$\mu$m, i.e. approximately one fourth of the
beam full width at half maximum (FWHM).

\vspace*{-2ex}
\paragraph{Photometry:}

Since the flux calibration of SPIRE is strictly valid for point sources only, we used
our empirical beams (corrected for the finite size of Uranus) to estimate beam areas
of 468, 850, and 1763 square arcseconds, respectively at 250, 350, and 500\,$\mu$m,
and obtain extended-source flux calibration. They are larger than the current
official estimates (based on Gaussian fits, i.e. not including sidelobes and spikes)
by 10 to 20\%.

We measure global flux densities of $(457. \pm 2.)$\,Jy at 250\,$\mu$m, $(155. \pm 2.)$\,Jy
at 350\,$\mu$m, and $(49.6 \pm 0.9)$\,Jy at 500\,$\mu$m (error bars not including
systematic flux calibration uncertainties of 15\%, nor beam size uncertainties).
They were multiplied by the recommended corrective factors \citep[see][]{Pohlen10}.

\vspace*{-2ex}
\paragraph{Source decomposition:}

\begin{figure*}
\centering
\vspace*{-1.5cm}
\hspace*{-1.8cm} \includegraphics[width=6.5cm]{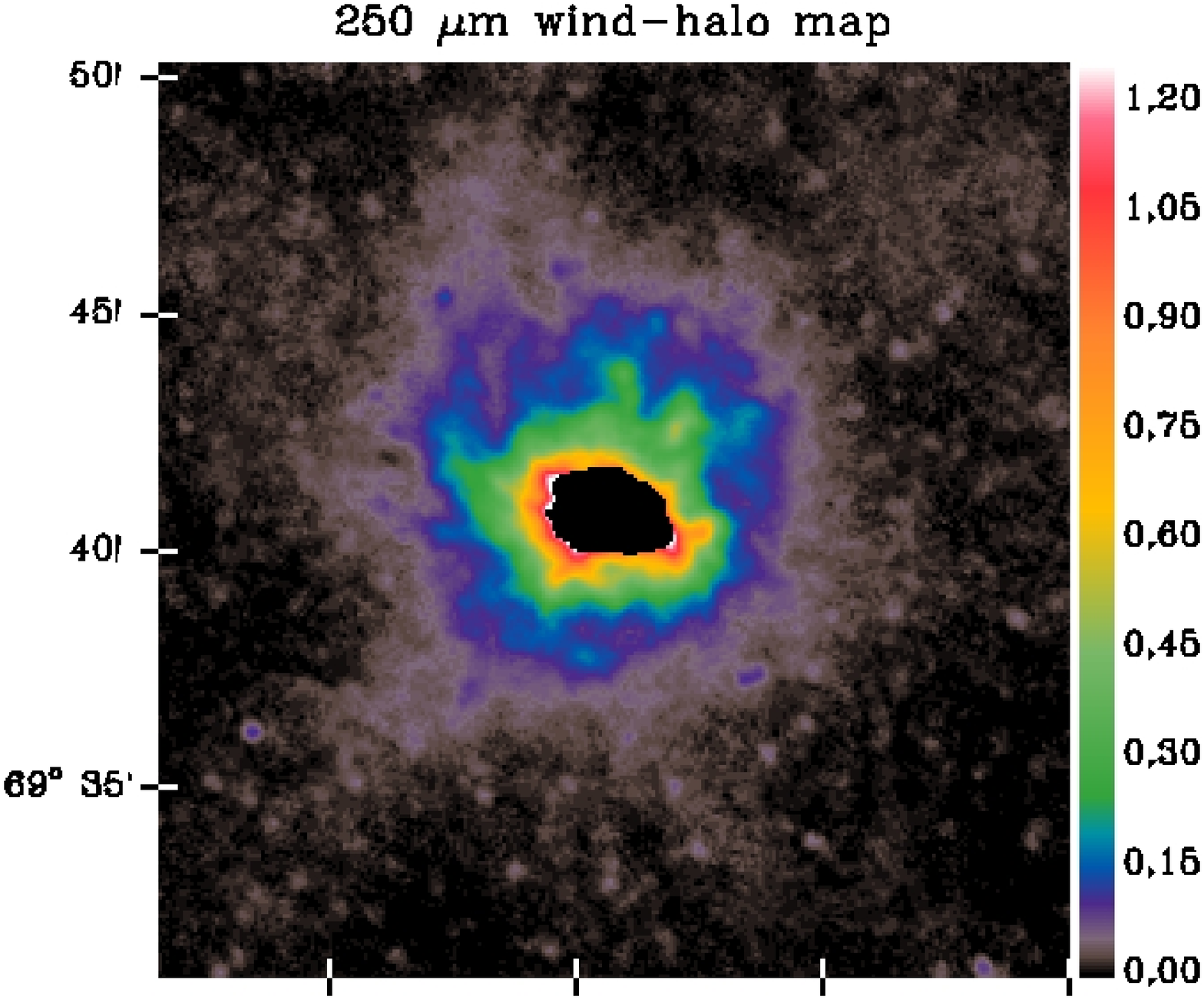}
                 \includegraphics[width=6.5cm]{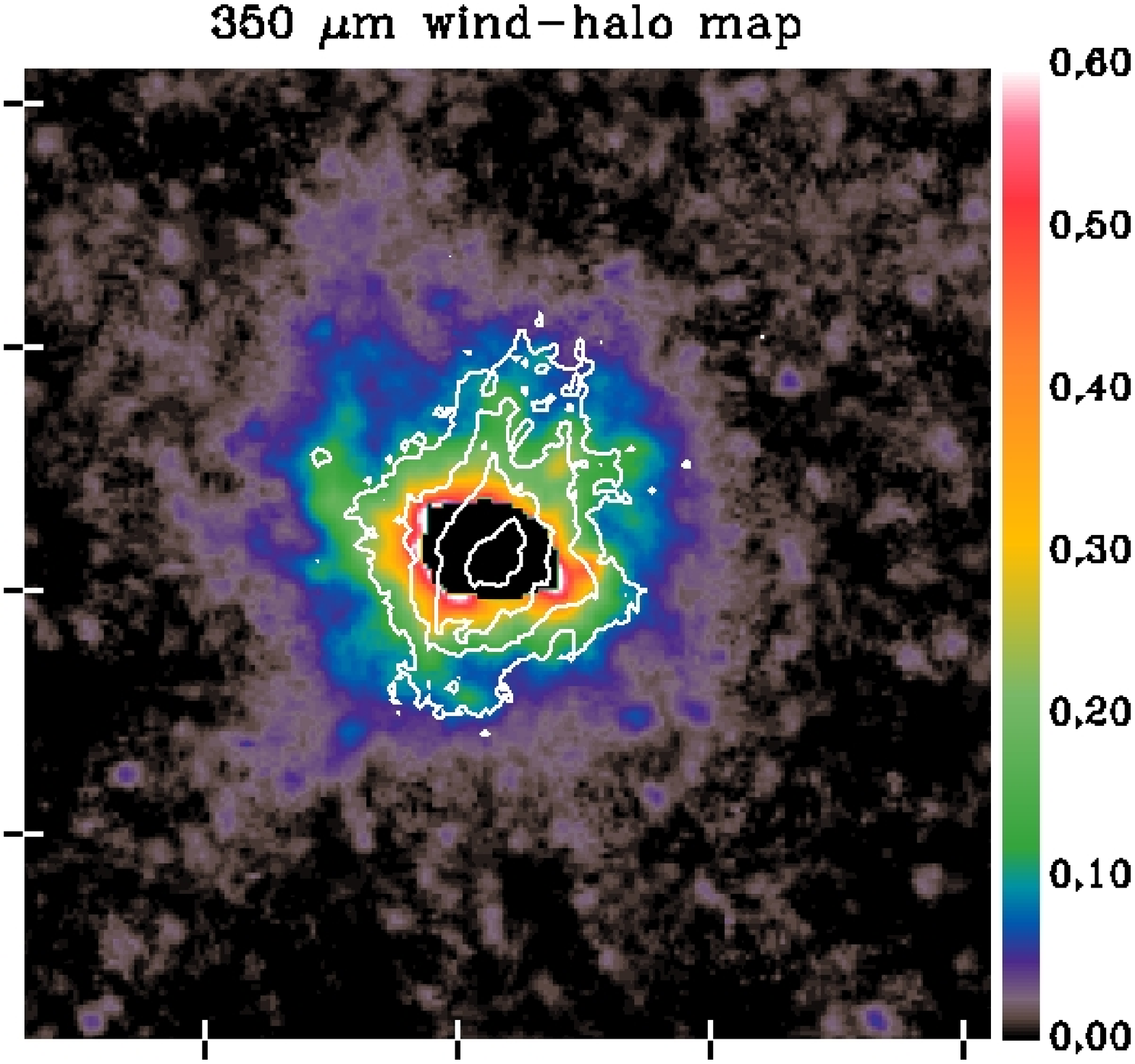}
\hspace*{-0.9cm} \includegraphics[width=6.5cm]{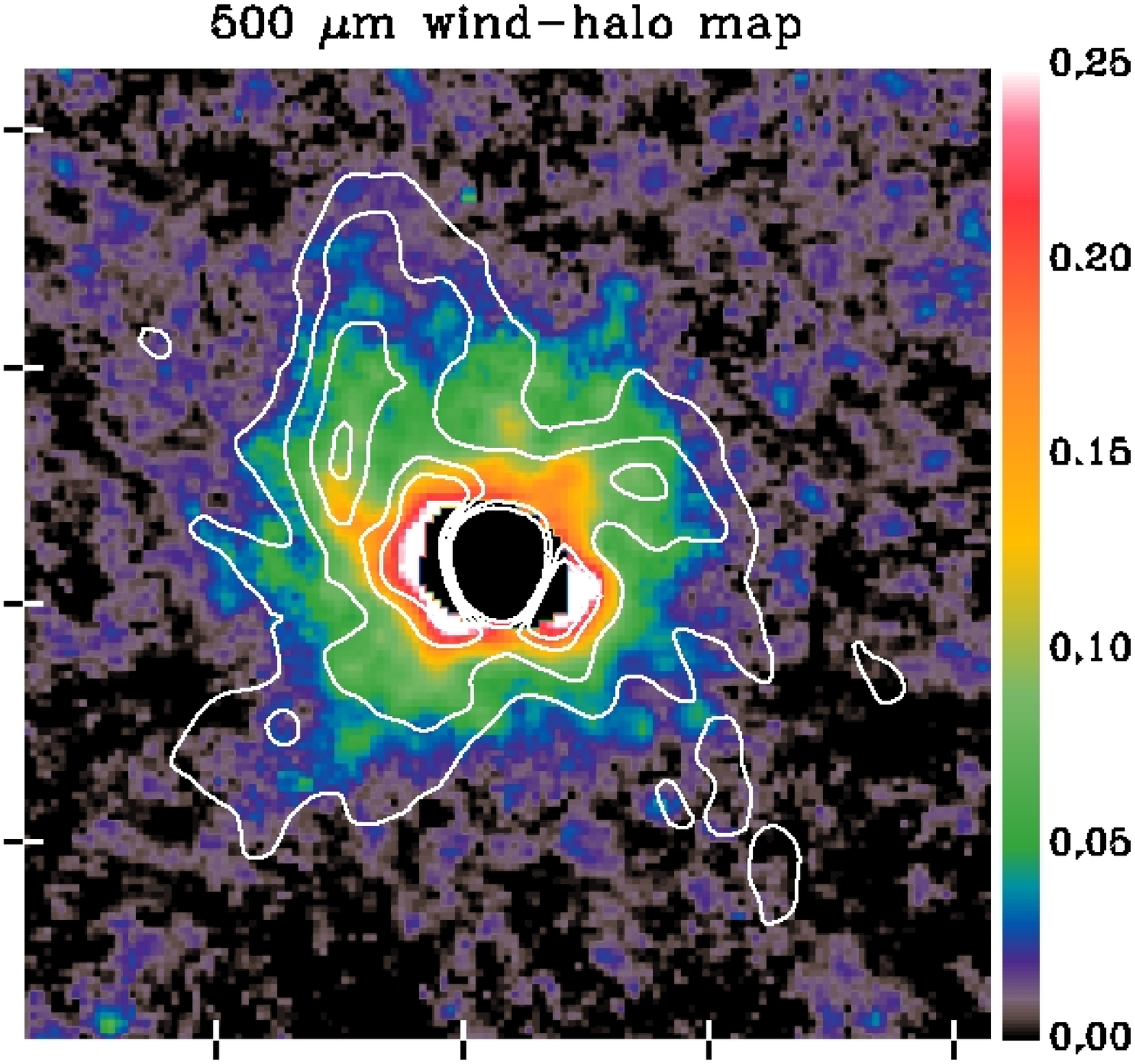}
\hspace*{-3cm} ~ \vspace*{-0.4cm} \\
\hspace*{-1.8cm} \includegraphics[width=6.5cm]{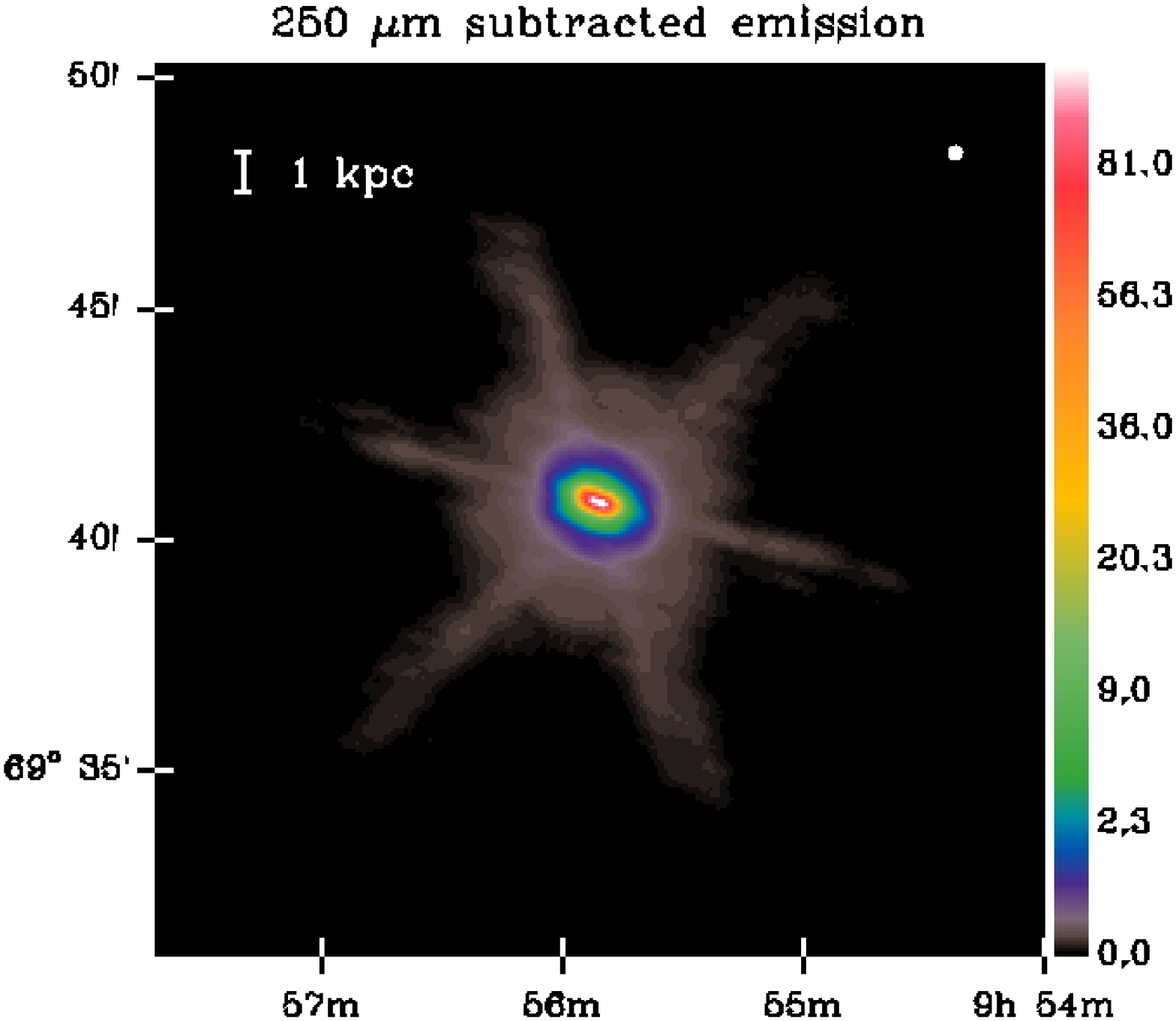}
                 \includegraphics[width=6.5cm]{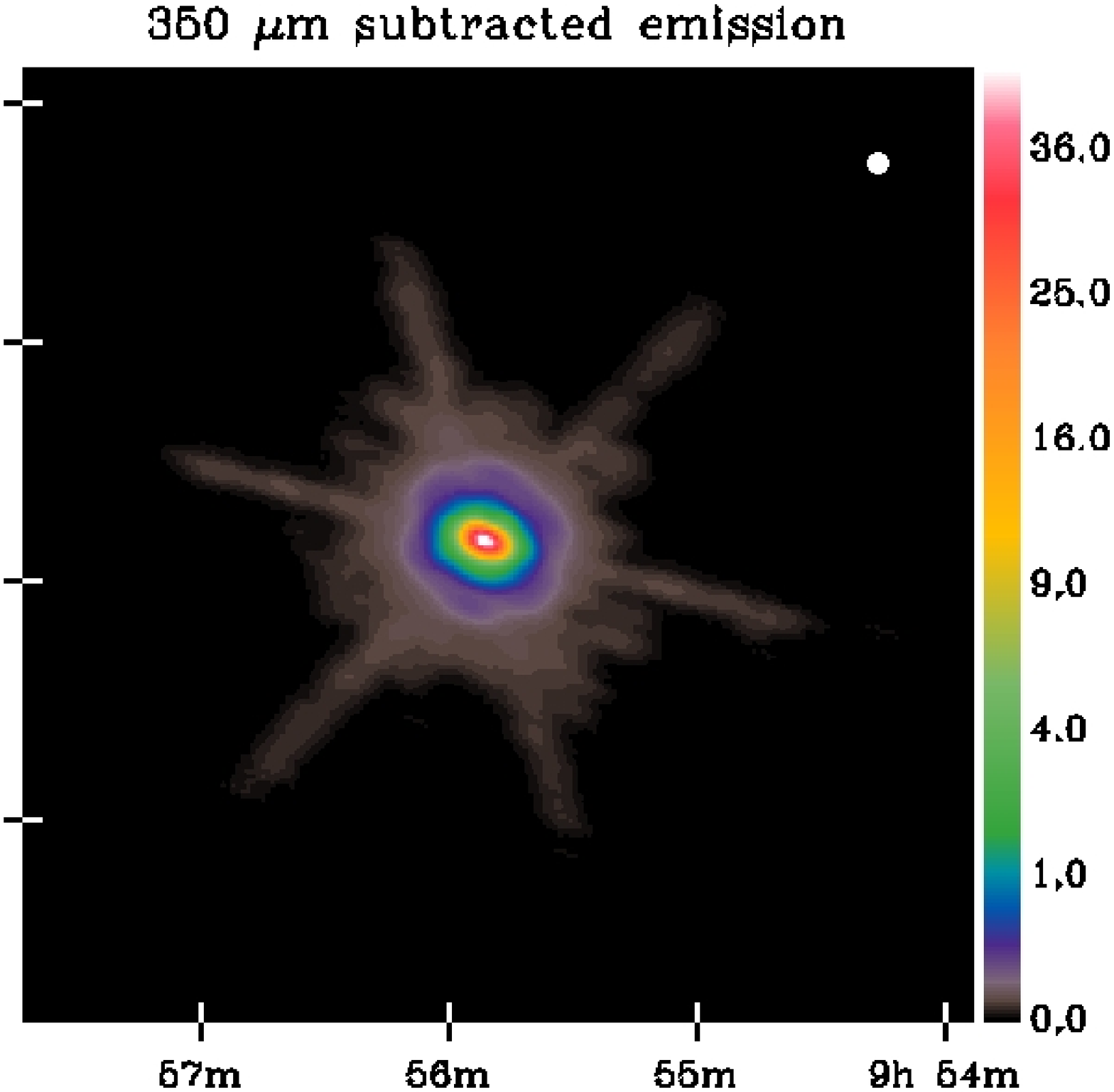}
\hspace*{-0.9cm} \includegraphics[width=6.5cm]{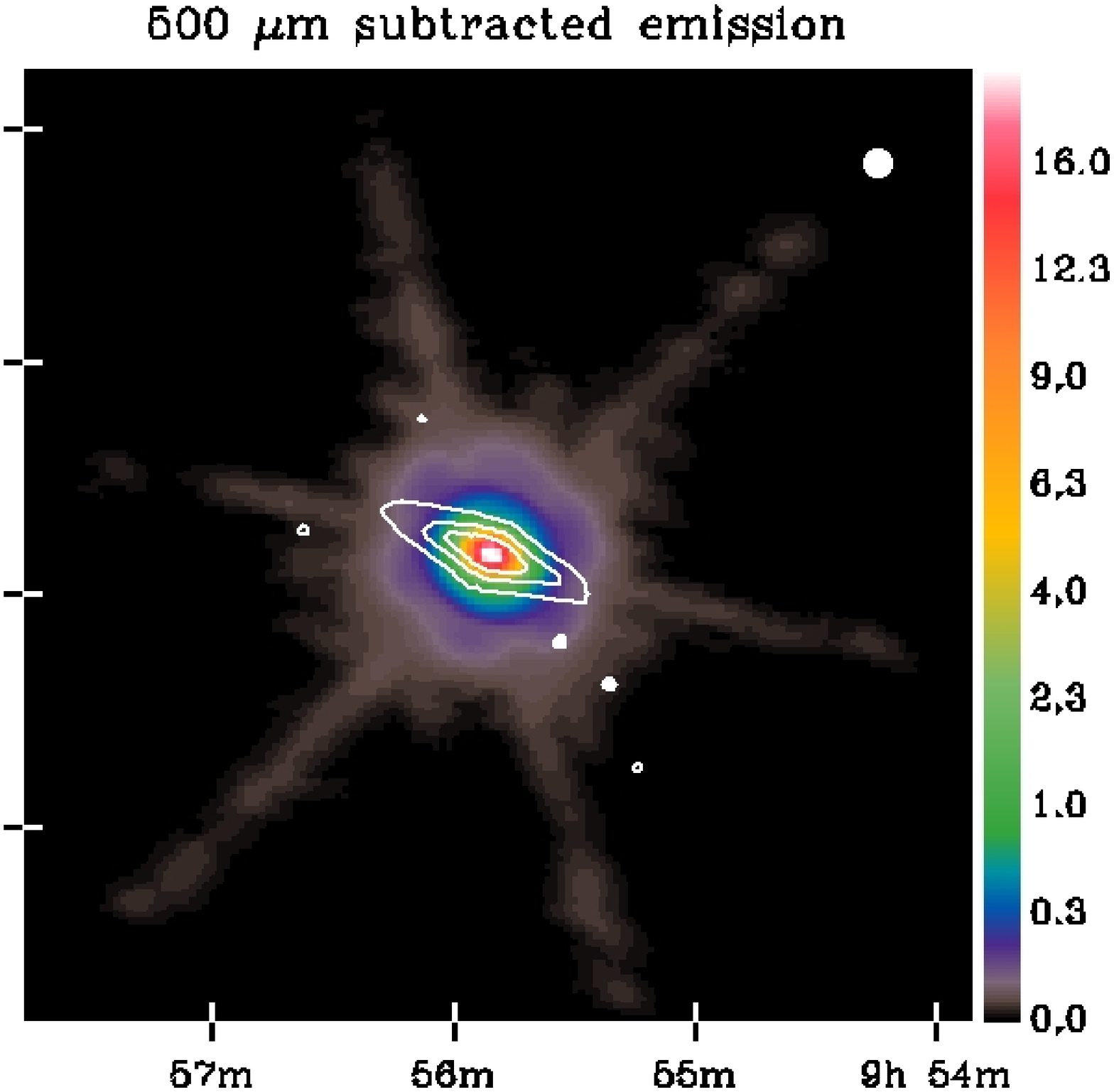}
\hspace*{-3cm} ~ \\
\caption{SPIRE maps resulting from the ``cleaning'' process. {\bf Top row:} residual maps
showing the wind and halo emission. Contours drawn from the HI map of \cite{Yun94} and from
the H$\alpha$-[NII] map of \cite{Boselli02} are superimposed on the 500\,$\mu$m and
350\,$\mu$m maps, respectively.
{\bf Bottom row:} subtracted central emission. The stellar disk is delineated by the
IRAC 3.6\,$\mu$m contours on the 500\,$\mu$m map. The initial maps have peak
brightnesses of 105, 48.2, and 21.6 Jy/beam, at 250, 350, and 500\,$\mu$m respectively.
The beam FWHM is sketched to the top right.}
\label{fig:cleaned}
\end{figure*}

The central regions of M\,82 have a complex structure. They contain multiple bright
infrared sources that can be most easily separated in mid-infrared images
\citep{Telesco92, Engelbracht06}.
The surface brightness dynamic range between the starburst and the halo
emission spans nearly three decades. Therefore, to subtract the central sources and
obtain maps of the superwind and halo as minimally contaminated by the diffraction
pattern of the bright center as possible, we adapted the {\small CLEAN} algorithm
used in radio astronomy \citep{Hogbom74} to substitute ``clean'' beams for beams with
extended sidelobes. At each iteration, the peak of the map is found, and the scaled
PRF at the same location is multiplied by a small gain and subtracted from the map;
a narrower Gaussian with the same flux is added to a model map. The residual maps
obtained after ``cleaning'' are shown in Fig.\,\ref{fig:cleaned}.

The starburst emission that we recover from the beam ``cleaning'' process, after adding
back to the model maps the underlying residuals (that account for about 7\% of the total),
amounts to 337, 111, and 35.4\,Jy at 250, 350, and 500\,$\mu$m, respectively, i.e. between
70\% and 75\% of the global emission. If we instead sum the emission from the four
brightest sources selected from the 350\,$\mu$m SCUBA map, fitted with fixed
positions and widths ($\sim 150$\,pc) in all maps, then we already account for about
57\% of the global emission in each SPIRE band.

Since there is a smooth transition from the disk to the off-disk emission, the spatial
separation between the two is somewhat arbitrary. We note that the disk of M\,82 is very
thick and turbulent, and that the residual maps are still contaminated by disk emission.
We thus masked pixels where the original flux density exceeds the cleaning threshold,
as shown in Fig.\,\ref{fig:cleaned}.

\vspace*{-2ex}
\paragraph{Dust opacities and emissivities:}

In all the following, we adopt the Galactic emissivity law derived by \cite{Reach95}
from the COBE survey. Dust grain opacities vary as a function of environment and
are very uncertain. However, dust masses can
easily be rescaled if another value is assumed. We thus adopt an opacity valid for dust
in the diffuse interstellar medium, $\kappa = 1.62$\,m$^2$\,kg$^{-1}$ in the DIRBE
140\,$\mu$m band \citep{Li01}. Converting it to an opacity at 500\,$\mu$m,
we obtain 0.18\,m$^2$\,kg$^{-1}$, in good agreement with what would have been obtained
from the 850\,$\mu$m opacity of \cite{James02}, i.e. $(0.20 \pm 0.06)$\,m$^2$\,kg$^{-1}$.

\section{The wind and tidal streams}

Cool dust is detected at $5 \sigma$ significance in filaments out to 9\,kpc from the center,
at 250\,$\mu$m. The maximal extensions are to the north-east and to the south-east of the
center, i.e. in filaments making a large angle with the polar axes of the galaxy and outflow.
As already remarked by \cite{Engelbracht06} for the PAH emission, dust occupies a very
extended halo, and is not confined to the edges of what is traditionally called the superwind.
Comparison with the HI map of \cite{Yun94} shows that the densest parts of the tidal
streams are associated with some of the extended dust detected in SPIRE maps. Below,
we provide quantitative arguments to support the impression that most of the dust mass
outside the disk of M\,82 has been removed by tidal interaction, rather than entrained
by the superwind.

\begin{figure*}
\centering
\includegraphics[width=8cm]{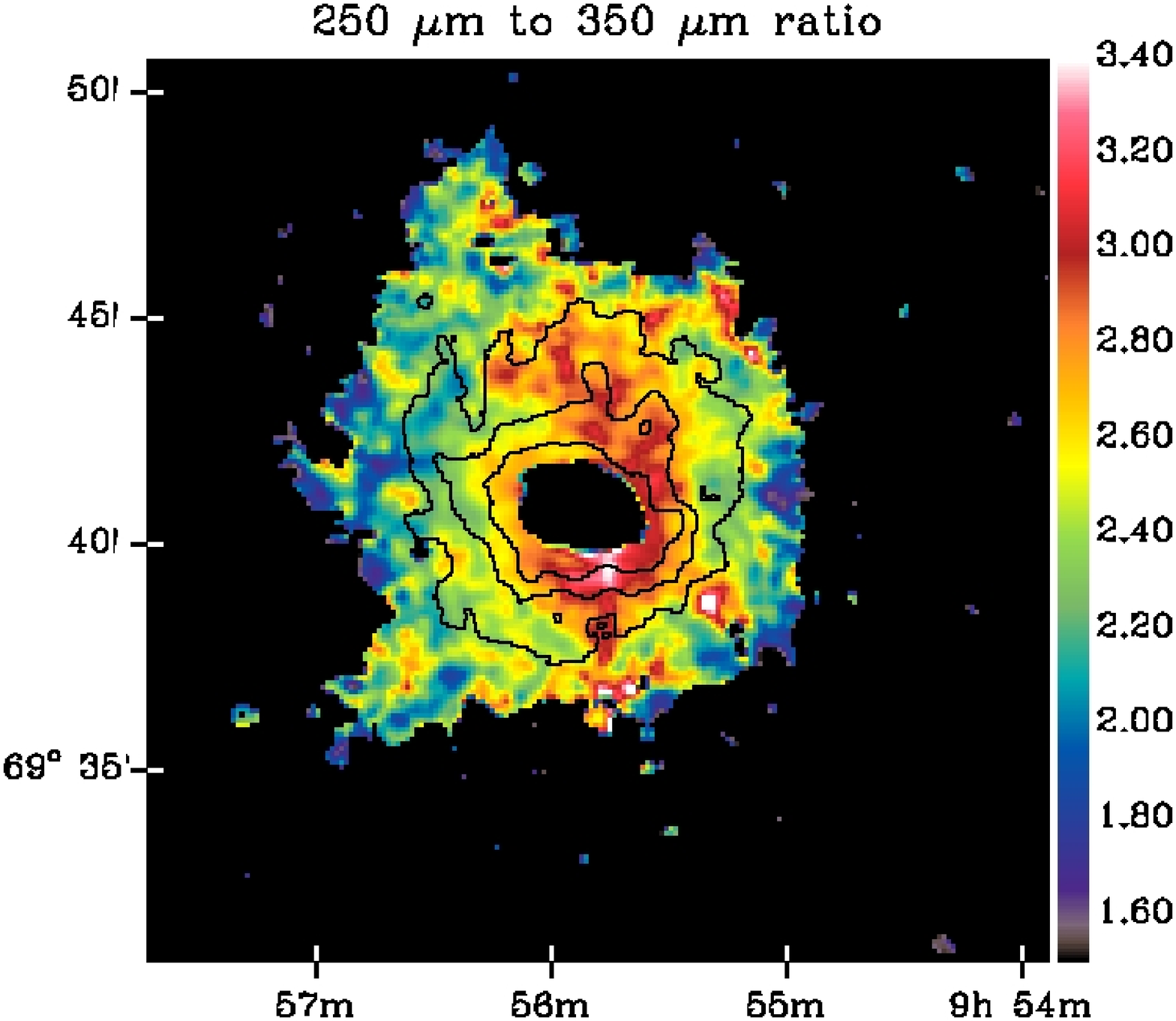}
\includegraphics[width=8cm]{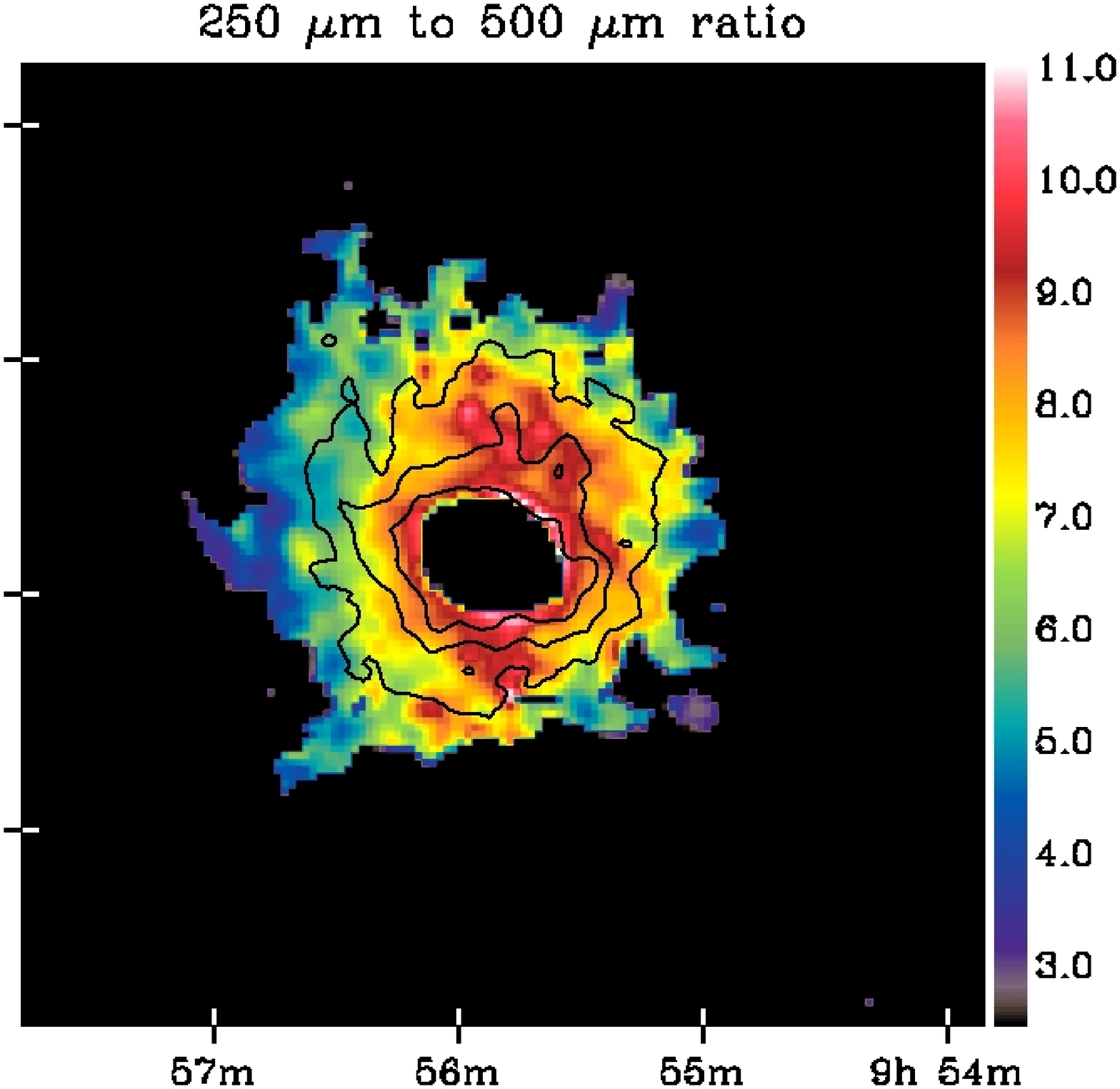}
\vspace*{-0.45cm} \\
\includegraphics[width=8cm]{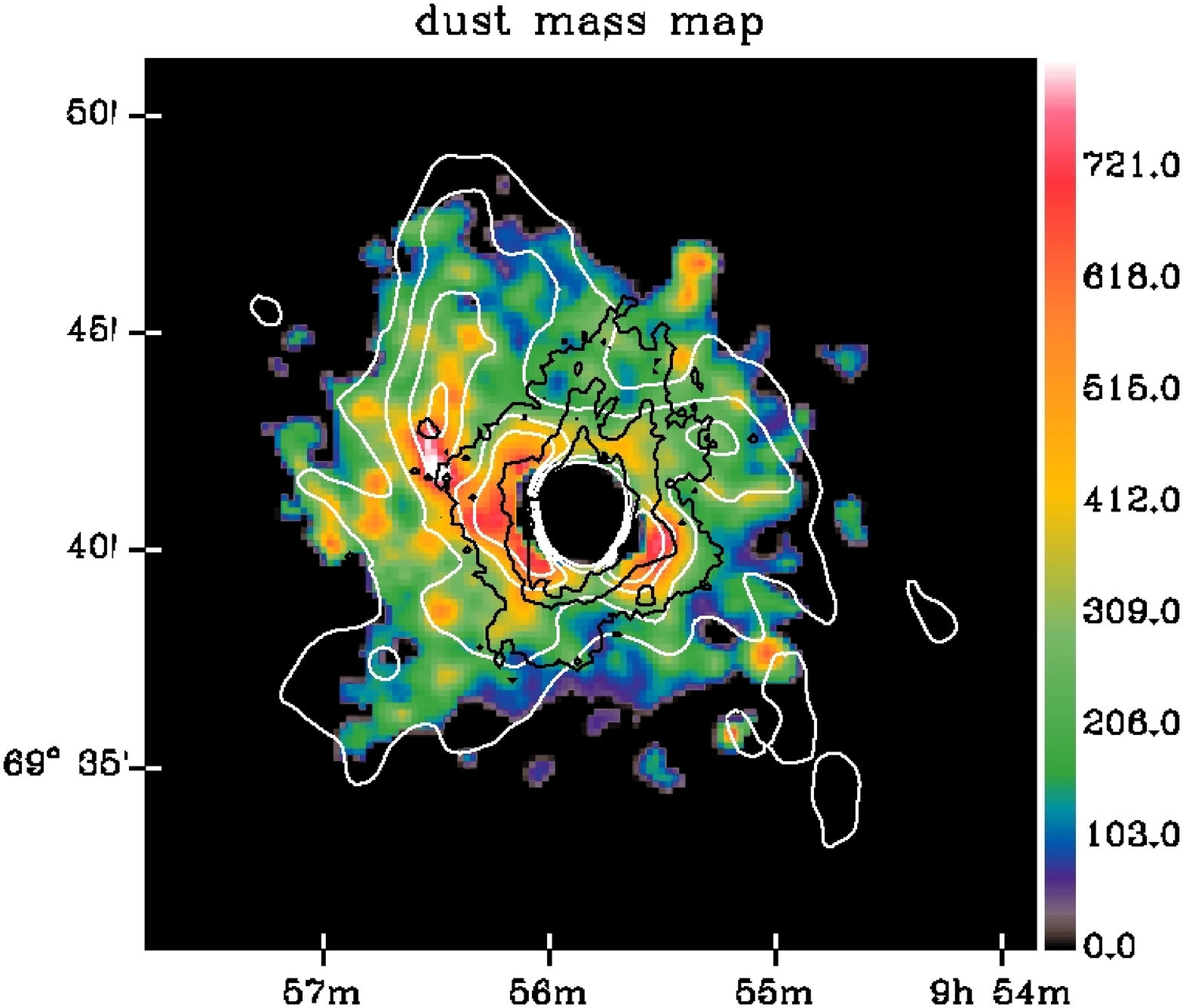}
\includegraphics[width=8cm]{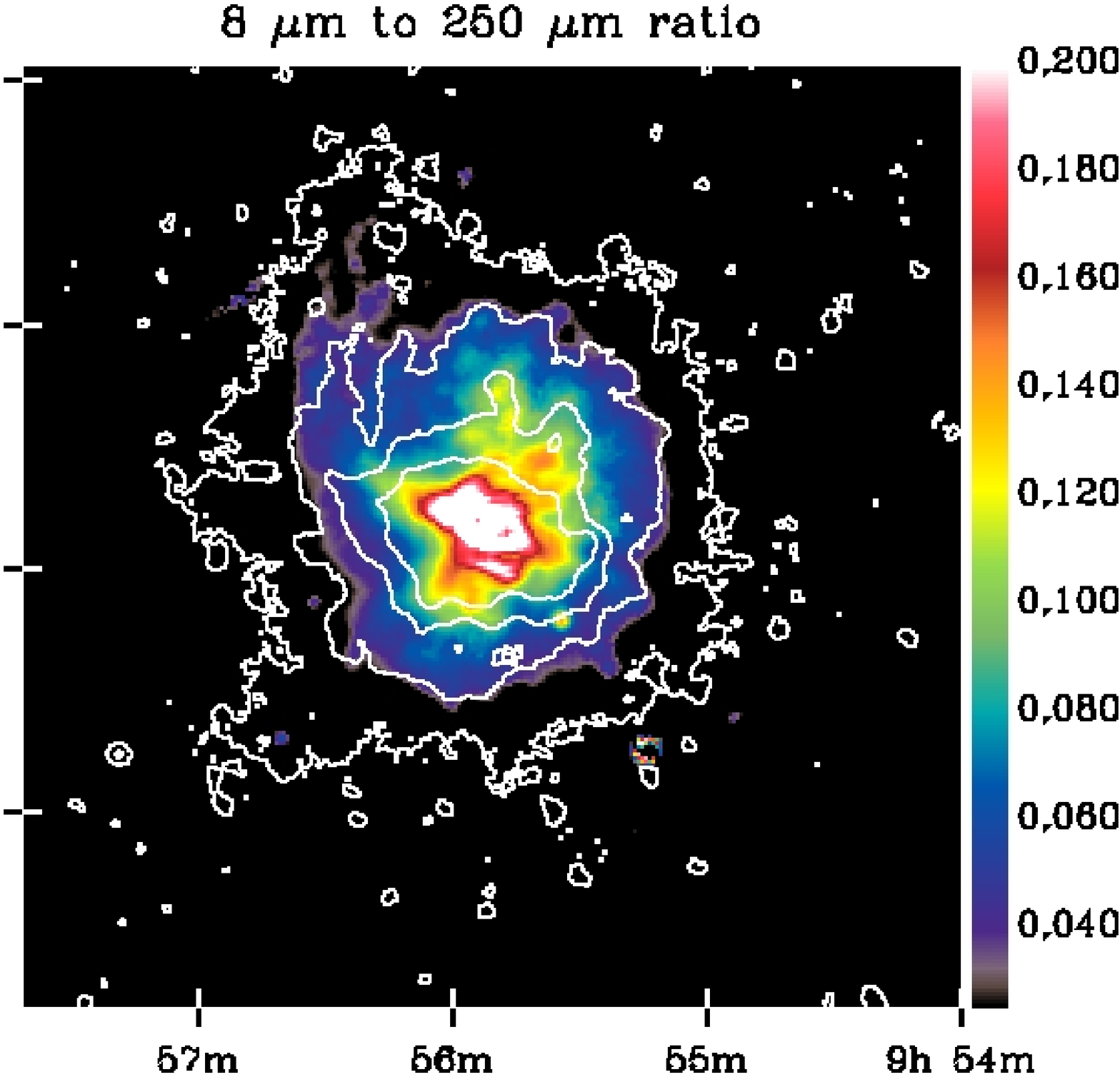}
\caption{{\bf Top row:} Flux ratio maps.
Left: 250\,$\mu$m to 350\,$\mu$m ratio, at 28$^{\prime\prime}$ angular resolution (FWHM).
Right: 250\,$\mu$m to 500\,$\mu$m ratio, at 40$^{\prime\prime}$ angular resolution (FWHM).
Contours of the 250\,$\mu$m map are overlaid in black.
{\bf Bottom row:} Left: dust mass map in M$_{\sun}$ per pixel (FWHM of 40$^{\prime\prime}$,
corresponding to 760\,pc; pixels are 170\,pc on a side).
HI contours are superimposed in white and H$\alpha$ contours in black.
{\it Special note:} the dust blob located around 09:55:23 +69:46:10 is likely associated
with M\,82, rather than a background source, since its morphology in all SPIRE bands
is diffuse. Right: 8\,$\mu$m to 250\,$\mu$m flux ratio map of the wind and halo,
at the 250\,$\mu$m resolution. Contours of the 250\,$\mu$m map are overlaid.
\vspace*{-5cm}
~}
\label{fig:wind_colors}
\end{figure*}

To approximately match the angular resolution of the maps and examine spatial
variations in dust colors of the wind and halo, we convolved the ``cleaned'' maps with
Gaussian kernels of appropriate widths, and regridded the resulting maps to the pixel
size at the longest wavelength. Figure\,\ref{fig:wind_colors} shows the 250\,$\mu$m to
350\,$\mu$m ratio map and the 250\,$\mu$m to 500\,$\mu$m ratio map. Based on these
ratios, the dust temperature in the wind and halo ranges between about 12\,K and 50\,K,
and its average value decreases with projected distance from the starburst.
Some filaments that are much warmer than their surroundings can be seen; they are located
in-between bright dust spurs (Fig.\,\ref{fig:wind_colors}), and thus must have relatively
low densities. The present data do not allow us to test whether these color variations
are caused purely by different excitation conditions, or whether they are affected by
changes in the dust composition within the wind. Assuming that these filaments follow
the Galactic emissivity law, we estimate that they have temperatures between 30\,K
and 50\,K, i.e. at least as high as in the starburst. They seem to be associated with
the starburst wind, rather than with the tidal features traced by the HI map.

We estimate the total dust mass in the wind and halo from the 500\,$\mu$m and dust
temperature maps, the latter derived from the 250\,$\mu$m to 500\,$\mu$m ratio.
We obtain $M_{\rm d~halo} = (1.1 \pm 0.5) \times 10^6$\,M$_{\sun}$ (mass error map
summed linearly, and $2.5 \sigma$ cut applied to the brightness maps). The mass derived
from the global halo fluxes and a single-temperature fit (with $T_{\rm dust} = 22.3$
to 25.6\,K) is similar: $M_{\rm d~halo} = (1.0$ to $1.2) \times 10^6$\,M$_{\sun}$.
This is much lower than the SCUBA estimate of \cite{Alton99}
(2 to $10 \times 10^6$\,M$_{\sun}$), even using the same dust opacity. We cannot rule
out a cold component undetected by SPIRE, but the discrepancy may predominantly
arise from several analysis issues: although the halo is barely detected in the
SCUBA maps (because they are only 2.6\,kpc in diameter, and the chop throw placed
the sky reference within the wind), the starburst and disk have not been subtracted
with their sidelobes, and the halo emission may thus be severely contaminated by them;
the procedure followed by \cite{Alton99} to estimate the dust temperature also
entails large errors.

The combination of the dust temperature and brightness maps shows that dust that is
spatially coincident with the superwind (likely to have been entrained by it) is warmer
and more diffuse than dust that is associated with the tidal streams, the latter
consequently dominant in terms of mass, as illustrated by the dust mass map
in Fig.\,\ref{fig:wind_colors}. Defining large sectors to exclude the superwind regions,
based on H$\alpha$ and X-ray maps (with opening angles of 74$^{\rm o}$ in the north
and 64$^{\rm o}$ in the south), we argue that at a minimum 65\% of the halo dust mass
is not associated with the superwind. Since tidal features overlap in projection with
superwind features, this fraction is a strict lower limit.

The global dust mass of M\,82, folding in the 1.25\,mm measurement of \cite{Thuma00},
is $M_{\rm d~tot} \sim (4.5 \pm 1.4) \times 10^6$\,M$_{\sun}$
(with $T_{\rm dust} = 24$\,K, which provides a fit too low by only 3\% at 350\,$\mu$m
and by 22\% at 250\,$\mu$m). If our mass estimates are correct, they imply that
$\sim 25$\% of all the dust has been expelled from the disk. Either the efficiency of
gas and dust ejection by the superwind and the tidal interaction is very high, or the
dust grain composition and size distribution in the wind and halo significantly
differ from those in the starburst. Without data of comparable quality at both shorter
and longer wavelengths, we cannot test the second alternative, but we deem it
less likely.

Since the interstellar medium of M\,82 must have become extremely turbulent and porous
as a result of the sustained injection of mechanical energy by the successive starbursts,
probably enough ultraviolet radiation is leaking out of the disk to provide
most of the dust heating necessary to account for the observed emission. GALEX images
show diffuse far-UV and near-UV light scattered by dust grains out to large distances
\citep{Hoopes05}, at least up to 6\,kpc on the southern side. However, this light is
mainly confined to lobes around the minor axis, whereas dust emission is not.
Optical photons must dominate the heating further away from the minor axis.

Using the archival Spitzer 8\,$\mu$m image to trace PAHs in the wind \citep{Engelbracht06},
we find that the PAH emission does not extend as far out as the 250-500\,$\mu$m emission,
and that there is a steep radial gradient in the 8\,$\mu$m to 250\,$\mu$m flux ratio
(Fig.\,\ref{fig:wind_colors}). There are also some regions of enhanced PAH brightness
relatively close to the polar axis, which correspond well to the 250\,$\mu$m spurs coincident
with H$\alpha$ emission. These regions contain relatively warm dust. We hypothesize that they
are well outside the wind shock, where PAHs would have very short lifetimes \citep{Micelotta10},
and that they simply correspond to more efficient illumination by UV photons leaking through
discrete cavities in the disk. A large fraction of optical line emission in this region
is also due to photoionization, and not only to shocks \citep{Shopbell98}.
The general gradient in the 8\,$\mu$m to 250\,$\mu$m flux ratio and in the far-infrared
flux ratios would then be a natural consequence of the dilution of the radiation field
with distance from the emitting stars.

\section{By way of conclusion}

The intergalactic medium can be enriched in metals by both tidal stripping and
starburst winds. M\,82 offers an exquisite example of both processes at work in the
same galaxy.
Considering the spatial distributions of the dust brightness and temperature in the
halo, the tidal interaction may be far more effective than the wind
in removing the cool interstellar medium from the disk of M\,82. We estimate that
more than two thirds of the extraplanar dust mass is associated with the tidal streams
traced by atomic hydrogen.
If dust is able to survive for a long time in the extragalactic medium around
such interacting systems, which may have been common in the past, then it will have
significant column densities along lines of sight to distant galaxies, and
cause extra attenuation outside the Milky Way.

\acknowledgements

We thank the referee for suggesting useful clarifications, L. Leeuw
and I. Robson for providing their SCUBA images.
SPIRE has been developed by a consortium of institutes led by Cardiff Univ. (UK):
Univ. Lethbridge (Canada); NAOC (China); CEA, LAM (France);
IFSI, Univ. Padua (Italy); IAC (Spain); Stockholm Obs. (Sweden);
Imperial College London, RAL, UCL-MSSL, UKATC, Univ. Sussex (UK);
and Caltech, JPL, NHSC, Univ. Colorado (USA).
This development has been supported by national funding agencies: CSA (Canada);
NAOC (China); CEA, CNES, CNRS (France); ASI (Italy); MCINN (Spain);
Stockholm Obs. (Sweden); STFC (UK); and NASA (USA).

\newpage

\end{document}